\documentclass[11]{article}%%%%%% to generate preprint number
\usepackage[margin=30mm]{geometry}
\usepackage{amsmath} %for dealing with mathematics,
\usepackage{amssymb}
\usepackage[pdftex]{graphicx} % for dealing with figures.
\usepackage{subfig} % for getting the subfigures e.g., "Figure 1a and 1b" etc.
\usepackage{authblk}

\title{Systematic effects on lensing reconstruction from a patchwork of CMB polarization maps}

\author[1]{Ryo Nagata}

\author[2]{Toshiya Namikawa}

\affil[1]{Japan Aerospace Exploration Agency (JAXA), Institute of Space and Astronautical Science (ISAS), Sagamihara, Kanagawa  252-5210, Japan 
}

\affil[2]{Kavli Institute for the Physics and Mathematics of the Universe (Kavli IPMU, WPI), UTIAS, The University of Tokyo, Kashiwa, Chiba 277-8583, Japan}

\begin{document}

\maketitle

\begin{abstract}%
We investigate the tolerance for systematic errors in lensing analysis applied to a patchwork map of Cosmic Microwave Background polarization. We focus on the properties of the individual polarization maps that comprise the patchwork and discuss the associated calibration residuals that are coherent on those subpatches. We numerically simulate the polarization field modulated as a whole patchwork and apply a suite of lensing analyses to reveal the response of the reconstructed gravitational lensing potential and delensing efficiency. At systematic error levels expected in the near future, we find that it is possible to accurately reconstruct the lensing potential on scales larger than the subpatch size and that there is no severe degradation of the lensing $B$-mode removal efficiency in the subsequent delensing analysis.
\end{abstract}

% \subjectindex{xxxx, xxx}

\section{Introduction} \label{introduction}
Signals arriving from sources at cosmological distances have their images modulated by gravitational lensing effects caused by density distribution present in their paths. The gravitational lensing effects in Cosmic Microwave Background (CMB) polarization maps have begun to be measured by observational experiments \cite{PB:phi:2019,P18:phi,BKXVII,SPT3G23,ACT24:phi} and promise to be a significant source of information for cosmological research \cite{LensRev2006,Namikawa2014b}. Observations of CMB polarization are a powerful clue to revealing the mass distribution of the lensing source. The large-scale mass distribution is a valuable source of information on the structural evolution of the universe and is also a promising analytical tool for extracting information on the inflationary universe from the CMB signals \cite{Seljak:2003pn,Smith:2010gu}.

The CMB polarization originating from density fluctuations forms even-parity patterns ($E$ modes) on the last scattering surface. The gravitational deflection of light rays due to the mass distribution in the light paths results in the secondary generation of odd parity patterns ($B$ modes) from the E-mode image, forming a signal floor of several $\mu$K-arcmin, which is called the lensing $B$ modes \cite{Zaldarriaga:1998:LensB}. In the next generation of CMB polarization observations, a sensitivity level of a few $\mu$K-arcmin will be realized by implementing densely integrated detector arrays, and it is expected to be possible to evaluate the multipole moments of the lensing $B$ modes as signal dominant data down to a few arcminutes' angular scale \cite{S42019}. On the other hand, there is a concern that the floor created by the lensing $B$ modes will provide a detection limit for inflationary primordial gravitational waves that improvements in experimental setup cannot overcome.

The primary $E$ modes on the last scattering surface are statistically isotropic on the celestial sphere. As a result of the lensing effect, the spatial pattern of the polarization is modulated, giving rise to the secondary $B$ modes, and the statistical property of the polarization map becomes anisotropic. The statistical anisotropy caused by the gravitational lensing effect can be used to extract information about the mass distribution along the light propagation path, which is the so-called gravitational lensing potential. Since the lensing potential directly captures the mass distribution, it provides an accurate map describing the density structure of the high-redshift universe. In particular, since the $B$ modes are expected to be dominated by signals originating from the lensing effects, the reconstruction analysis using the correlation between the $E$ and $B$ modes of CMB polarization has the potency to estimate the lensing potential down to about ten arcminutes with a high signal-to-noise ratio \cite{Hu2002}.

So far, several CMB observation projects have reconstructed the lensing potential, including POLARBEAR \cite{PB:phi:2019}, Planck \cite{P18:phi}, BICEP/{\it Keck} \cite{BKXVII}, SPT-3G \cite{SPT3G23}, and ACT \cite{ACT24:phi}. Future CMB observations, such as Simons Observatory \cite{SO2019} and CMB-S4 \cite{S42019}, will significantly improve sensitivity to the lensing potential. Since high resolution as well as high sensitivity are necessary to reconstruct the lensing potential with high accuracy, ground-based CMB telescopes with large antenna diameters will have a cost advantage. The total signal-to-noise ratio of the lensing potential reconstructed by CMB-S4 with its arcminute angular resolution is expected to reach several thousand. The reconstructed lensing potential can be used to remove the lensing $B$ modes and enhance the detectability of the primordial gravitational wave signals originating from the inflation in the early universe. Such analysis is called delensing. By using CMB-S4 level data, we can remove most of the lensing $B$-mode power.

In the current state of CMB polarization observations, space and ground-based observations are divided to search different angular domains \cite{Hazumi2020}. Although high-resolution polarization maps are provided by large ground-based CMB telescopes, low-frequency noise from atmospheric fluctuations and temporal variations in instrumental conditions limit the size of the region where coherent observations can be realized. Therefore, evaluating the lensing potential on dozen-degree angular scales has been challenging using conventional analysis methods. In our previous study \cite{Namikawa2014a}, we investigated a method to make a wide-area map of the lensing potential by stitching many observation patches of CMB polarization to realize the extraction of information on the lensing potential on large scales. In that study, we found that if the lensing reconstruction analysis is applied to the patchwork of the polarization maps without any treatment, the reconstructed lensing potential will be biased toward large angular scales. We applied a treatment that excludes information at scales larger than the size of individual patches from the analysis. We found that the reconstructed potential is accurate without that bias over a scale equivalent to the entire patchwork.

In this paper, we work on a wide-area reconstruction of the lensing potential using CMB polarization. We simulate polarization maps by assuming multiple experiments that observe different sky regions separately and reconstruct the lensing potential from a patchwork of the polarization maps obtained from the simulations. Sensitivity levels expected for the next generation of CMB polarization observations will allow us to measure the gravitational lensing effect with a high signal-to-noise ratio, and the limiting factor for the accuracy of the reconstruction will be systematic errors originating from instrumental characteristics and calibration uncertainties. In particular, in this study, we focus on the effect of systematic errors due to differences in the relative properties of the constituent polarization maps when reconstructing the lensing potential from the patchwork polarization map. In our situation, the entire observation domain is divided into small regions with mutually no coherence in absolute calibration errors. We simulate the effect on the patchwork polarization map of mutual deviations of absolute calibration residuals between the individual constituent patches and evaluate their impact on the estimation of the lensing potential. In the past, we developed a framework for evaluating how systematic errors propagate to the results of the lensing analysis \cite{SYSPTEP2021}. In that framework, a series of processes were formulated, from applying systematic errors to the polarization maps, lensing reconstruction, delensing, and the identification of mean-field bias by a Monte Carlo method. This study applies the framework to the absolute calibration residuals associated with the patchwork polarization map.

The paper is organized as follows: In Section \ref{mapsim}, we describe our simulation procedure for applying systematic errors to the patchwork polarization map. In Sections \ref{reconstruction} and \ref{delensing}, we present the results of the lensing reconstruction analysis and the subsequent delensing analysis. Finally, Section \ref{summary} provides some discussion and conclusions.

\section{Map simulation} \label{mapsim}
In this section, we describe the synthesis of the patchwork polarization map and the application of systematic errors in our numerical simulations. In our simulations, we consistently use the $\Lambda$CDM parameter set based on the baseline likelihood analysis of Planck 2018 \cite{Planck2018}. The specific parameter values are $h = 0.6736$, $\Omega_b h^2 = 0.02237$, $\Omega_c h^2 = 0.12$, $\Omega_{\nu} h^2 = 0.000645$, $n_s = 0.965$, $A_s(k_{\rm pivot}=0.05{\rm Mpc}^{-1})=2.099\times10^{-9}$, and $\tau = 0.0544$. One neutrino species with mass and two (effectively $2.046$) without mass are assumed. Since the signal amplitude of the primordial gravitational waves (i.e., the tensor-to-scalar ratio $r$) has only the upper bound, $r < 0.036 \, (2\sigma)$ \cite{BKXIII,BKXV}, we set it vanishing. In order to clarify how errors in the absolute calibration of the individual constituent polarization maps affect the lensing analysis, we performed our simulations with a setup in which its polarization maps are assumed to be noiseless, and there is no image blurring due to finite beam width. We used the publicly available software, {\tt Lenspix} \footnote{https://cosmologist.info/lenspix/}, to produce a lensed CMB polarization map.

\subsection{Sky partition} \label{mapsim_sky}
We prepare a patchwork map consisting of small individual polarization maps (henceforth referred to as subpatches) that assume regions observed separately in ground-based experiments. We use this patchwork map, which as a whole covers a large region of the sky, as the target of our lensing reconstruction. Initially, the map has spatially coherent CMB information over a large region of the sky. By applying independent systematic error realizations in the regions of individual subpatches, as described in the following subsection, each subpatch has the characteristics of separately observed data. In this paper, we consider a simple case where the total area of the patchwork map corresponds to the entire sky. The geometry of the subpatches that make up the patchwork follows the {\tt HEALPix} \footnote{https://healpix.jpl.nasa.gov/} pixelization scheme, and the size of each subpatch is set to be about $7$ degrees (corresponding to $\ell \sim 50$ in the power spectrum domain).

\subsection{Systematic errors} \label{mapsim_sys}

\begin{figure*}
\begin{center}
\includegraphics{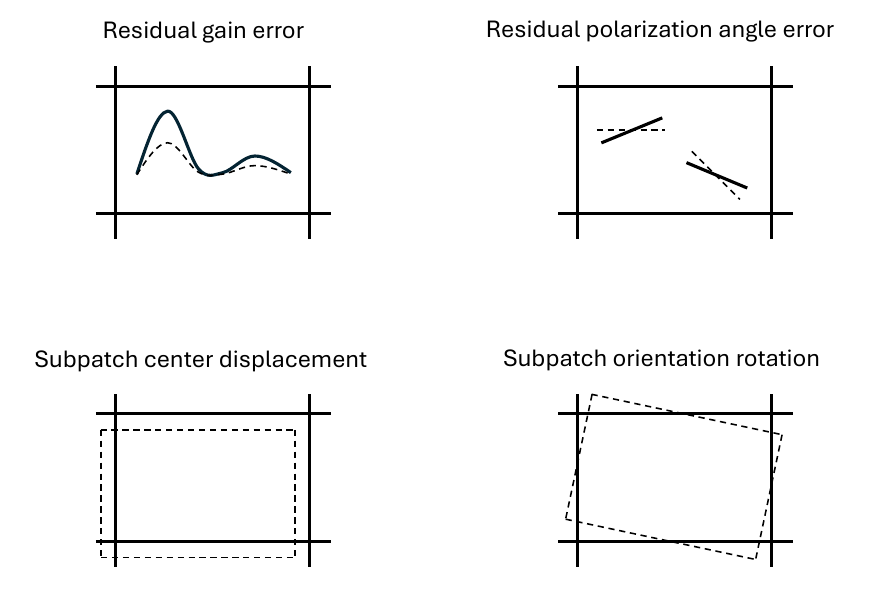}
\caption{Illustrations of the four systematic errors. The dashed lines indicate the original signal amplitude, polarization orientations, image position, or image orientation.}
\label{sys_illust}
\end{center}
\end{figure*}

It is assumed that the polarization maps within their associated subpatches are made at different times and under different observing conditions, sometimes provided by different projects with different telescopes. Therefore, calibration levels for observational parameters such as signal gain, polarization angle, or pointing direction are expected to vary from subpatch to subpatch. If we consider, for example, a uniform gain error in a subpatch, it is easy to see that it does not produce false $B$ modes when considered only in the subpatch. Absolute calibration errors in the entire subpatch do not necessarily affect the lensing analysis when the subpatch is treated alone. However, relative differences between subpatches in the correction residuals of those errors add modulation on scales larger than the subpatch size in the entire patchwork, resulting in systematic errors that depend on the degree of the modulation. Our concern is that those systematic errors propagate through the lensing analysis to the reconstructed lensing potential, introducing bias and raising uncertainty in the reconstruction.

Errors in gain, polarization angle, and pointing are well-known systematic errors frequently discussed in CMB polarization observations. This study investigates gain, polarization angle, subpatch center displacement, and subpatch orientation rotation, which are described in the following paragraphs and explained graphically in Figure \ref{sys_illust}. Writing a typical error magnitude as $\sigma$, we determine the actual error value for each subpatch from random numbers following the Gaussian distribution $N(0,\sigma^2)$. In the cases of subpatch misalignments, the direction of the misalignment is determined from random numbers following a uniform distribution. To apply those systematic errors to the polarization maps, we used routines of the {\tt Healpix} library. \\

Residual gain error: Gain is the correspondence between incident radiation and readout electrical signal. Variations caused by various instrumental and environmental factors during observation result in correction residuals in the reference absolute gain levels. Such uncertainty is expressed by multiplying a polarization map by a numerical factor. In our simulations, the factor is uniform within each subpatch and takes different values between subpatches. \\

Residual polarization angle error: Uncertainty of polarization angle originates from various disturbances and instrument calibration errors and propagates to post-observation analysis. Although errors in reference directions of polarization angle can be corrected in individual subpatches with the aid of $EB$ correlation \cite{EBcalib2020}, the correction residual for each subpatch still results in non-uniform modulation of the patchwork as a whole. Such polarization angle error is expressed by multiplying a polarization map by a complex phase factor, which is implemented in the simulation in much the same way as in the gain case. Note that this is the only exception among the four types of systematic errors treated in this paper, which causes conversion between $E$ and $B$ modes even within a subpatch. \\

Subpatch center displacement: Locations of subpatches on the celestial sphere are subject to errors depending on uncertainties in pointing models and telescopes' pointing calibration. For each subpatch, displacement of the subpatch center from where it should be causes a modulation pattern larger than the subpatch size in the entire patchwork polarization map. This modulation not only leads to leakage from $E$ modes to $B$ modes (and vice versa) but also to missing information in parts of the sky. In the implementation of our numerical simulation, we use uniformly distributed random numbers in the interval from $0$ to $360$ degrees to determine the direction of the displacement and rotate the entire sky so that the subpatch center moves along the great circle in that direction. Then, the original subpatch area contains a polarization map, which was initially a short distance away from the original subpatch area. This operation is applied to each subpatch. (Due to this operation, some images on the celestial sphere are lost. Therefore, remapping with a vector field cannot strictly represent this operation.) \\

Subpatch orientation rotation: We refer to the deviation of orientations of subpatches from their original directions (while the subpatches' centers retain their positions in the sky) as subpatch orientation rotation. Subpatch orientation rotation, like subpatch center displacement, is caused by pointing errors. In the implementation of our numerical simulation, we apply a rotation operation to the entire sky, with the center of a subpatch concerned as its rotation axis, and repeat this procedure for each subpatch. The direction of the rotation is randomly chosen to be clockwise or counterclockwise with equal probability. Also in this case, images located at the edges of the subpatch are partially lost. \\

Before moving on to the discussion of the lensing analysis, we give an overview of those modulations' behaviors in the power spectrum domain. In the cases of gain error and polarization angle error, the modulation pattern is represented by a scalar function. The angular spectrum has a nearly constant amplitude on the large scale side (up to $\ell \sim 50$ corresponding to the subpatch size) and rings down towards the small scale side with a power of about $\ell^{-3}$, forming troughs with intervals of $\Delta \ell \sim 50$. We can interpret this behavior as roughly equivalent to that of the spherical harmonic transform of a top hat function. Also, in the cases of subpatch misalignments, we can evaluate the angular spectra of their modulation patterns by the spin-1 spherical harmonic transformation by expressing positional misalignment at each sky point as a vector. This representation is not exact because image loss, which is caused by the actual operations, is not included. Still, it allows us to estimate the approximate behavior of the modulation pattern in the form of the angular spectrum. The behavior of their spectra is similar to those of the gain and polarization angle cases. Exceptionally, in the case of subpatch orientation rotation, the amplitude is slightly suppressed on scales larger than the subpatch size. This behavior comes from the fact that the distribution of the misalignment vectors caused by the subpatch orientation rotations is unlikely to have a larger structure than the subpatch size.

\section{Lensing reconstruction} \label{reconstruction}
In this section, we discuss the lensing reconstruction analysis from the patchwork polarization map with the systematic errors introduced in the previous section. Lensing reconstruction by $EB$ correlation plays a significant role in high-sensitivity, high-resolution CMB polarization observations we are considering here. The following equation gives the $EB$ quadratic estimator \cite{Okamoto2003}:

\begin{equation}\label{EB-est}
\begin{split}
[\widehat{\phi}^{EB}_{LM}]^* = A^{EB}_L\sum_{\ell\ell'}\sum_{mm'}
 \left(
  \begin{array}{ccc}
   \ell & \ell' & L \\
   m & m' & M \\
  \end{array}
 \right) 
g^{EB}_{\ell\ell'L}\widehat{E}_{\ell m}\widehat{B}_{\ell'm'} .
\end{split}
\end{equation}
Symbols with \quad $\widehat{}$ \quad represent quantities based on observed data. Symbols without \quad $\widehat{}$ \quad represent quantities theoretically determined prior to the lensing analysis. $\widehat{E}_{\ell m}$ and $\widehat{B}_{\ell'm'}$ are the harmonic expansion coefficients for the observed $E$ and $B$ modes. Here, $g^{EB}_{\ell\ell'L}$ and $A^{EB}_L$ are defined as
\begin{equation}\label{EB-g-A}
\begin{split}
g^{EB}_{\ell\ell'L} &= \frac{[f^{EB}_{\ell\ell'L}]^*}{C_{\ell}^{EE}C_{\ell'}^{BB}} , \\ 
A^{EB}_L &= \left\{ \frac{1}{2L+1}\sum_{\ell\ell'}f^{EB}_{\ell\ell'L}g^{EB}_{\ell\ell'L} \right\}^{-1} .
\end{split}
\end{equation}
$C_{\ell}^{EE}$ and $C_{\ell}^{BB}$ are the angular power spectra of $E$ and $B$ modes, respectively, including lensing effects \cite{Lewis2011,Anderes2013}. We assume these spectra do not include systematics' contributions. Although normalizations of the $E$ and $B$ modes can be adjusted after inspecting actual data, it does not suppress reconstruction errors in the situations concerned here. The weight function $f$ in the equations is given by 
\begin{equation}\label{EB-f}
\begin{split}
f^{EB}_{\ell\ell'L} = i C_{\ell}^{EE} S_{\ell' \ell L}^{(-)}, 
\end{split}
\end{equation}
where the mode coupling function $S_{\ell' \ell L}^{(-)}$ is defined as
\begin{equation}\label{EB-S}
\begin{split}
 S^{(-)}_{\ell\ell'L} &= \frac{1- (-1)^{\ell+\ell'+L}}{2} \sqrt{\frac{(2\ell+1)(2\ell'+1)(2L+1)}{16\pi}} \\ 
 &\qquad \times [-\ell(\ell+1)+\ell'(\ell'+1)+L(L+1)]
 \left(
  \begin{array}{ccc}
   \ell & \ell' & L \\
   2 & -2 & 0 \\
  \end{array}
 \right) .
\end{split}
\end{equation}
The angular spectrum of the reconstructed lensing potential ($\widehat{C}_{L}^{\phi\phi}$), which will appear in the following, is evaluated as the mean square of $\widehat{\phi}^{EB}_{LM}$ concerning $M$. This spectrum contains the Gaussian bias associated with the reconstruction. 

We use {\tt CAMB} \footnote{https://camb.info/} to compute the theoretical angular spectra. We use {\tt CmbRecTool} \footnote{https://github.com/toshiyan/} for lensing reconstruction and also for numerical calculation of the delensing template discussed in the next section. 

Our previous study shows that multipoles beyond the subpatch size need to be excluded from the reconstruction analysis to remove the bias in the reconstructed lensing potential \cite{Namikawa2014a}. We limit the multipoles to be included in the estimator evaluation to the range $\ell = 300 - 2048$. The lensing analysis that we address in this study also includes delensing. By this multipole limitation, we also avoid the well-known delensing bias \cite{Namikawa2014a,Lizancos:2021:delens-bias}.

When some spatial modulation is applied to a CMB polarization field, its reconstructed lensing potential is known to be biased, and the bias field therein is called a mean field \cite{Namikawa:2012:bhe}. When a modulation due to a systematic error is present, the estimator described above becomes biased, and the angular power spectrum includes the contribution of its associated mean field. To estimate the mean field for a given systematic error, we repeat our simulation procedure 200 times to obtain Monte Carlo samples and compute the sample mean of the estimators, which is regarded as the mean field. (The Monte Carlo samples are also used to evaluate the Gaussian bias by cross-correlating $E$ modes in its subset with $B$ modes in another subset.)

In the following four subsections, we present the result of our reconstruction analysis for each systematic error. For each error, we present two typical cases relating to its magnitude. In one of them, we are just beginning to see signs of the mean field in the reconstructed lensing potential. In the other, the feature of the mean field, including its spectral shape, is clearly recognizable. In the figures of systematics-induced $E$ and $B$ modes, the angular power spectra are evaluated by applying the usual spherical harmonic transformation without any special treatment for the patchwork.

The lensing reconstruction analysis is a method to estimate the lensing potential as a realization. In the figures in the following subsections (and also in Section \ref{delensing}), we show the analysis results of a realization. When we discuss trends in those results, they are based on shared properties among a few hundred simulation samples. 

\subsection{Residual gain error} \label{rec_gain}

\begin{figure*}
\hspace{-13pt}
\subfloat[]{
\includegraphics[width=3.05in]{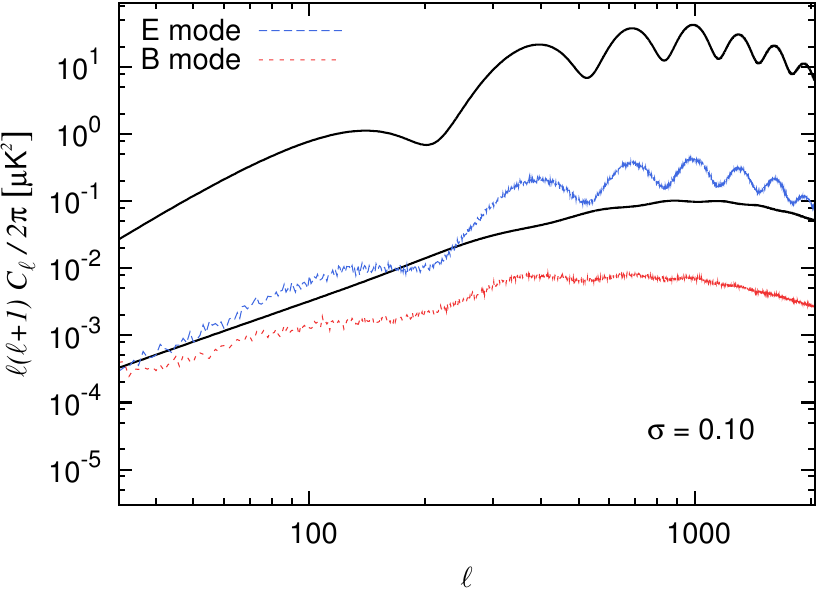}\label{clbb_sys_gain_s} }
\hspace{-4pt}
\subfloat[]{ 
\includegraphics[width=3.05in]{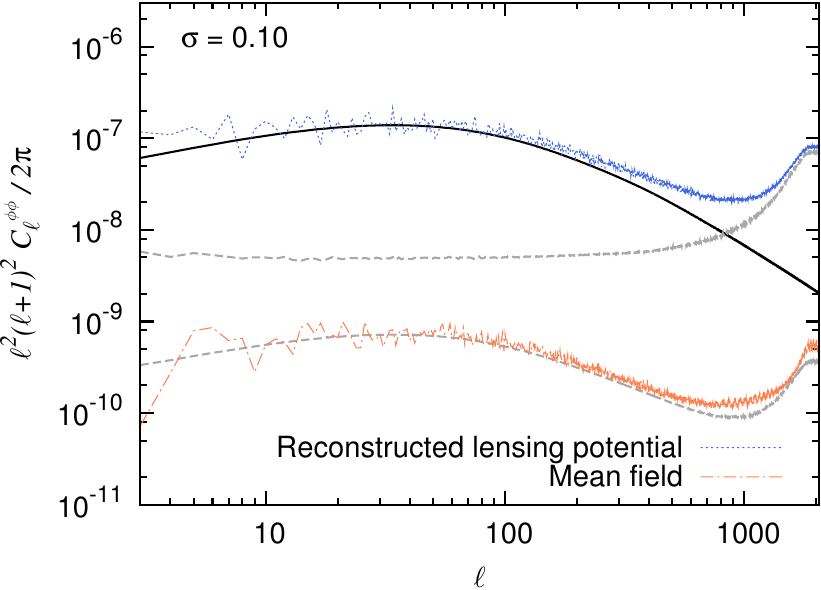}\label{clpp_mf_gain_s} }
\caption{Simulated angular power spectra for $\sigma = 0.10$ gain error. {\it Left}: angular power spectrum of systematics-induced $E$ modes (dashed blue) and that of induced $B$ modes (dotted red). The thick black curves are the theoretical CMB polarization spectra. {\it Right}: angular power spectrum of reconstructed lensing potential (dotted blue) and that of its Gaussian bias (upper dashed gray). The spectrum of the associated mean field (dot-dashed orange) and that of its Monte Carlo error (lower dashed gray) are also shown. The thick black curve is the theoretical lensing potential spectrum.}
\label{gain_s}
\end{figure*}

\begin{figure*}
\subfloat[]{
\hspace{-13pt}
\includegraphics[width=3.05in]{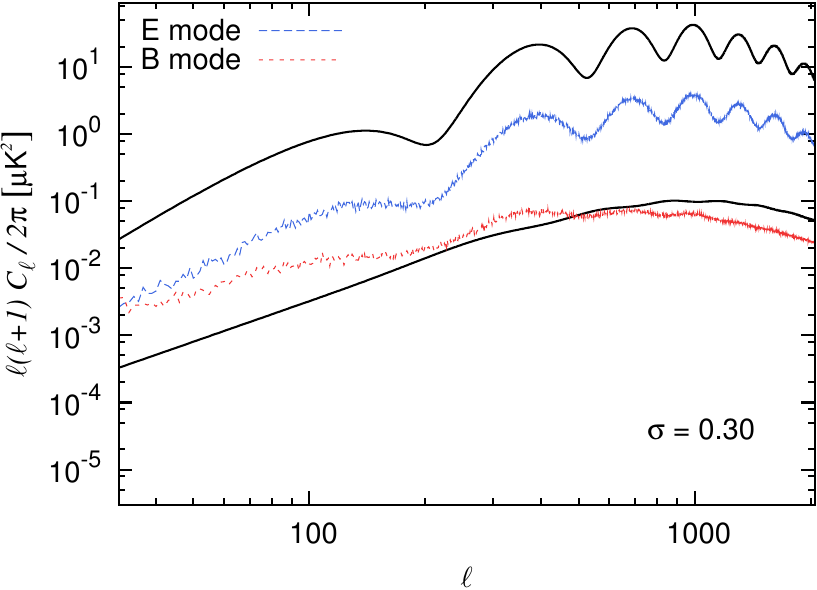}\label{clbb_sys_gain_l} }
\hspace{-4pt}
\subfloat[]{
\includegraphics[width=3.05in]{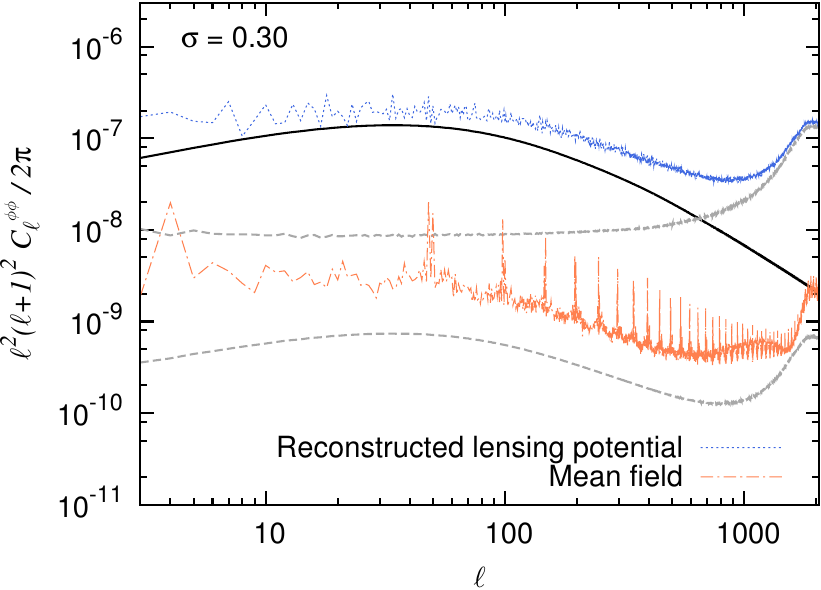}\label{clpp_mf_gain_l} }
\caption{Simulated angular power spectra for $\sigma = 0.30$ gain error. The line styles match the caption of Figure \ref{gain_s}.}
\label{gain_l}
\end{figure*}

Figures \ref{clbb_sys_gain_s} and \ref{clbb_sys_gain_l} show the angular spectra of the difference between the original CMB polarization map and the patchwork map with its applied gain errors in the subpatches. Here, the error magnitude parameters, $\sigma$, are given as fractional errors to actual gain. The modulation imposed on the patchwork map considered in this paper has the characteristic of being uniform within each subpatch, and the modulation-induced $E$ and $B$ modes show behavior that reflects it. In the case of gain error, the spectrum of $E$ modes originating from the error has the same shape as the original $E$-mode spectrum on the small scales. This comes from the fact that the gain errors merely give overall amplitude shifts inside the subpatches. On the other hand, the spatial distribution of polarization is modulated non-uniformly throughout the patchwork, which disturbs the even-parity property that the original $E$ modes have. As a result, a bias is induced in $B$ modes and it has a different spectral shape from that of the original $B$ modes.

The amplitude of the Gaussian bias in the reconstructed lensing potential responds to the magnitude of the $E$ and $B$ modes in the estimator. In the situation considered in this study, the induced $E$ and $B$ modes have very different amplitudes due to the error's coherence at the subpatch size. For the gain error, the impact of the induced $B$ modes on the overall $B$-mode signals is more significant than that of the induced $E$ modes on the overall $E$-mode signals, as seen in Figure \ref{clbb_sys_gain_l}. Therefore, it can be interpreted as the Gaussian bias associated with the reconstructed lensing potential increases as the induced $B$ modes bias the amplitude of the overall $B$ modes. (Compare Figure \ref{gain_s} with Figure \ref{gain_l}.) The mean-field bias of the reconstructed lensing potential has a structure corresponding to the subpatch size, which is observed in Figure \ref{clpp_mf_gain_l} as a sequence of spikes with the intervals of $\Delta \ell \sim 50$, which corresponds to the intervals between the troughs in the modulation pattern spectrum mentioned in Section \ref{mapsim_sys}.

The parameterization used here, the absolute gain errors of $0.1$ and $0.3$, is for demonstration purposes. The errors are far smaller in practical observations. Calibration errors of absolute gain have already reached a few percent, see, e.g., Ref. \cite{PB2017}.

\subsection{Residual polarization angle error}
\begin{figure*}
\hspace{-13pt}
\subfloat[]{
\includegraphics[width=3.05in]{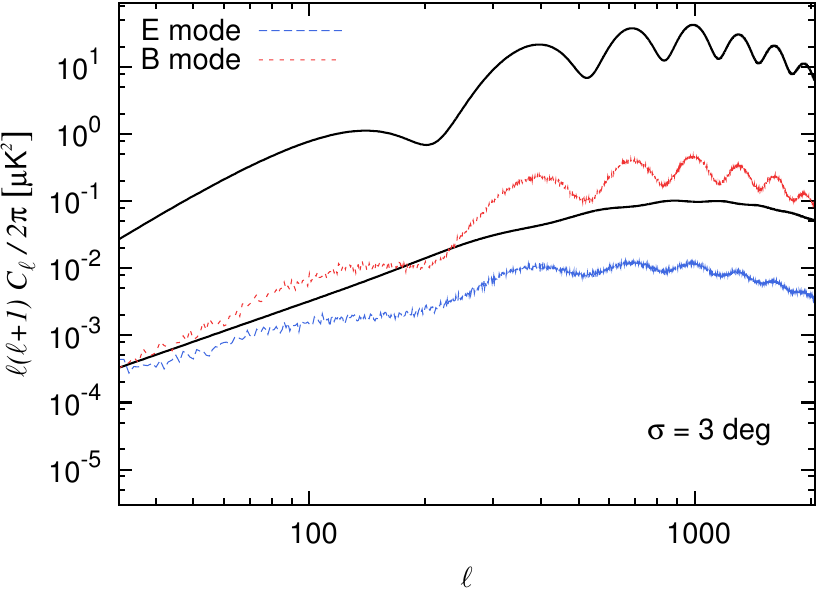}\label{clbb_sys_angle_s} }
\hspace{-4pt}
\subfloat[]{
\includegraphics[width=3.05in]{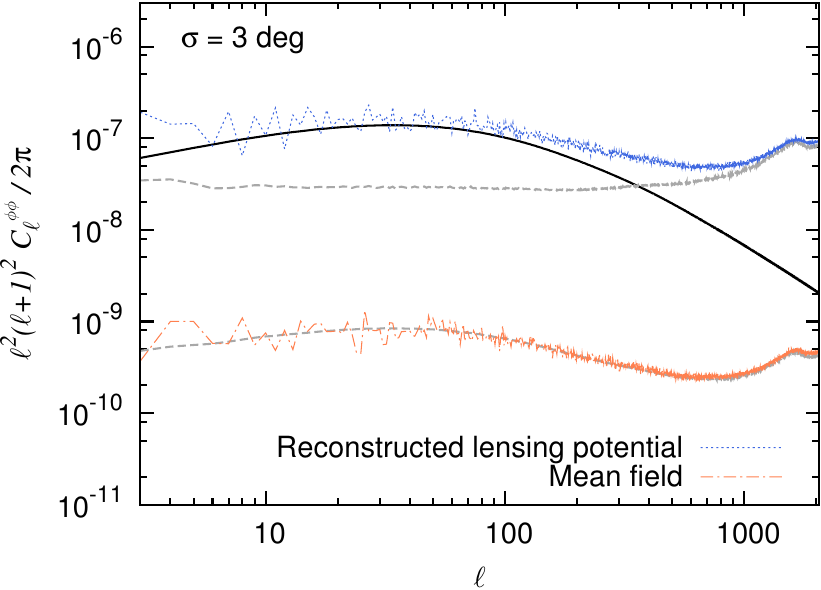}\label{clpp_mf_angle_s} }
\caption{Simulated angular power spectra for $\sigma = 3$deg polarization angle error. The line styles match the caption of Figure \ref{gain_s}.}
\label{angle_s}
\end{figure*}

\begin{figure*}
\hspace{-13pt}
\subfloat[]{
\includegraphics[width=3.05in]{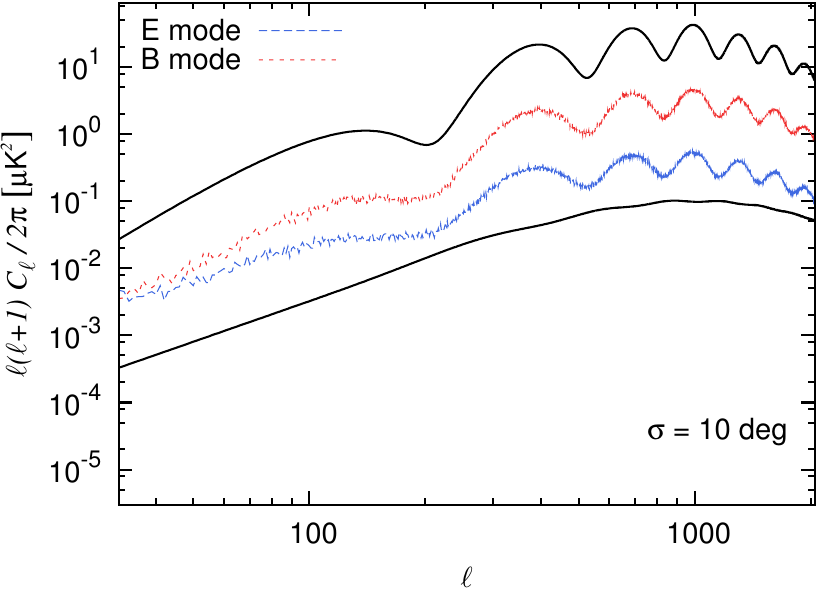}\label{clbb_sys_angle_l} }
\hspace{-4pt}
\subfloat[]{
\includegraphics[width=3.05in]{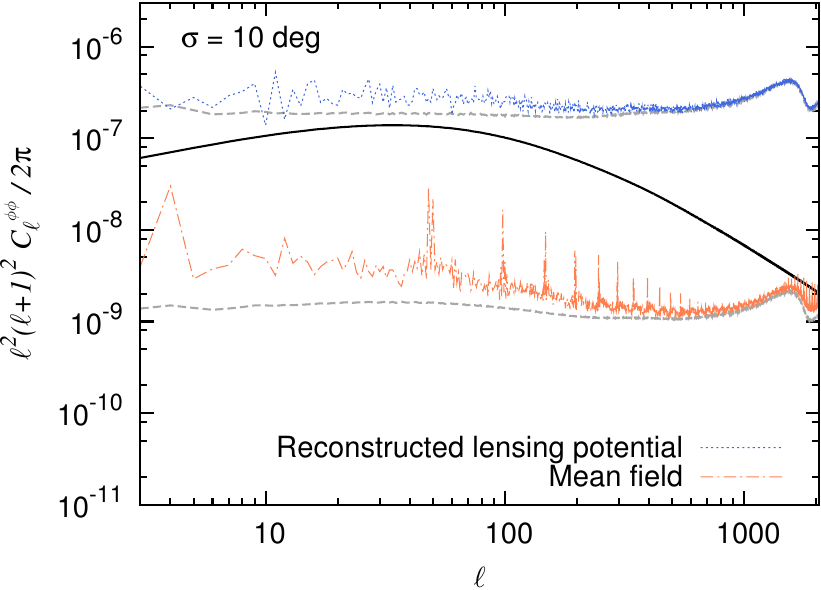}\label{clpp_mf_angle_l} }
\caption{Simulated angular power spectra for $\sigma = 10$deg polarization angle error. The line styles match the caption of Figure \ref{gain_s}.}
\label{angle_l}
\end{figure*}

Figures \ref{clbb_sys_angle_s} and \ref{clbb_sys_angle_l} show the spectra of the difference between the original CMB polarization map and the patchwork map with its applied angle errors in the subpatches. Since polarization orientations are uniformly rotated inside each subpatch, we can understand the results by the well-known leakage formula in literature, e.g., Ref. \cite{Keating2013}. Specifically, the $B$-mode spectrum induced by the angle error is approximately the original $E$-mode spectrum scaled by the square of the rotation angle. On the other hand, over somewhat large rotation angles, the induced $E$-mode spectrum roughly scales with the fourth power of the rotation angle. Again, since the impact on the cosmological CMB signals is more prominent in the $B$ modes, the Gaussian bias of the reconstructed lensing potential increases with the degree to which the induced $B$ modes bias the overall $B$ modes. In the case of polarization angle error, the response of the mean field to the imposed error is much milder than that of the Gaussian bias, as shown in Figures \ref{angle_s} and \ref{angle_l}.

In the next generation of CMB polarization observations, angle calibration errors are expected well within a few dozen arcminutes, e.g., Ref. \cite{EBcalib2020}. Judging from Figure \ref{angle_s}, the impacts on the Gaussian bias and mean field are small enough in future observations.

\subsection{Subpatch center displacement} \label{rec_disp}

\begin{figure*}
\hspace{-13pt}
\subfloat[]{
\includegraphics[width=3.05in]{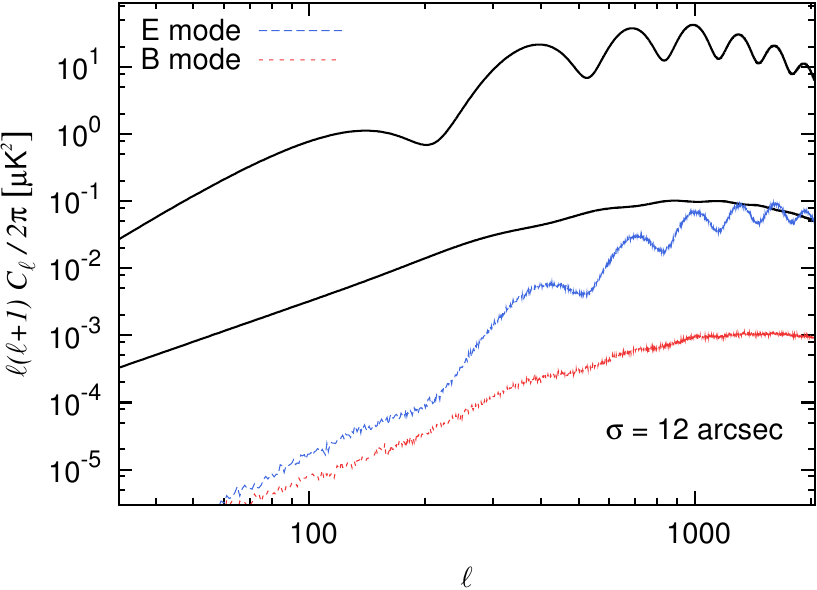}\label{clbb_sys_shift_s} }
\hspace{-4pt}
\subfloat[]{
\includegraphics[width=3.05in]{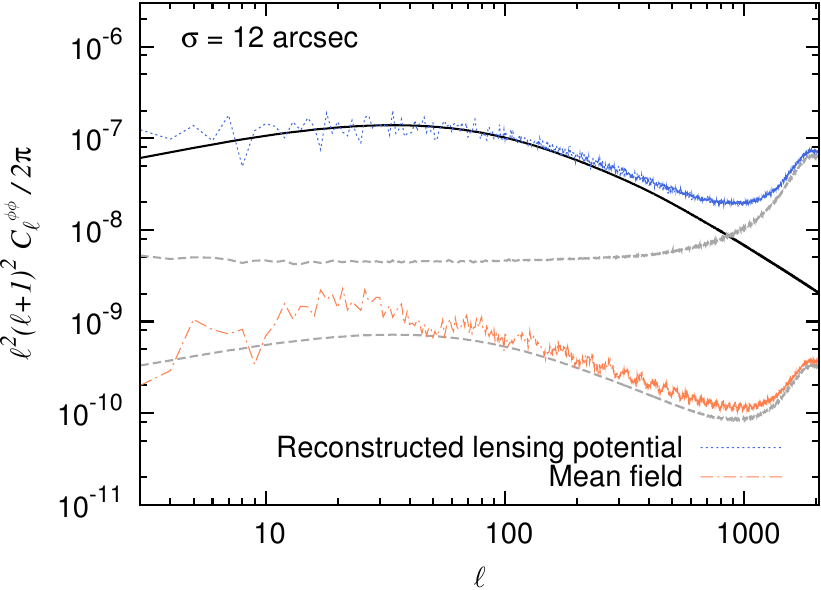}\label{clpp_mf_shift_s} }
\caption{Simulated angular power spectra for $\sigma = 12$arcsec subpatch center displacement. The line styles match the caption of Figure \ref{gain_s}.}
\label{shift_s}
\end{figure*}

\begin{figure*}
\hspace{-13pt}
\subfloat[]{
\includegraphics[width=3.05in]{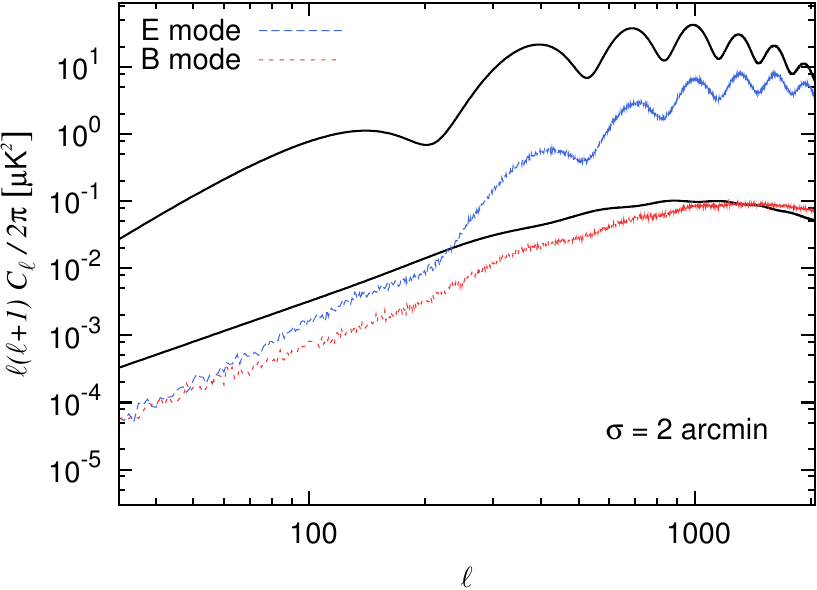}\label{clbb_sys_shift_l} }
\hspace{-4pt}
\subfloat[]{
\includegraphics[width=3.05in]{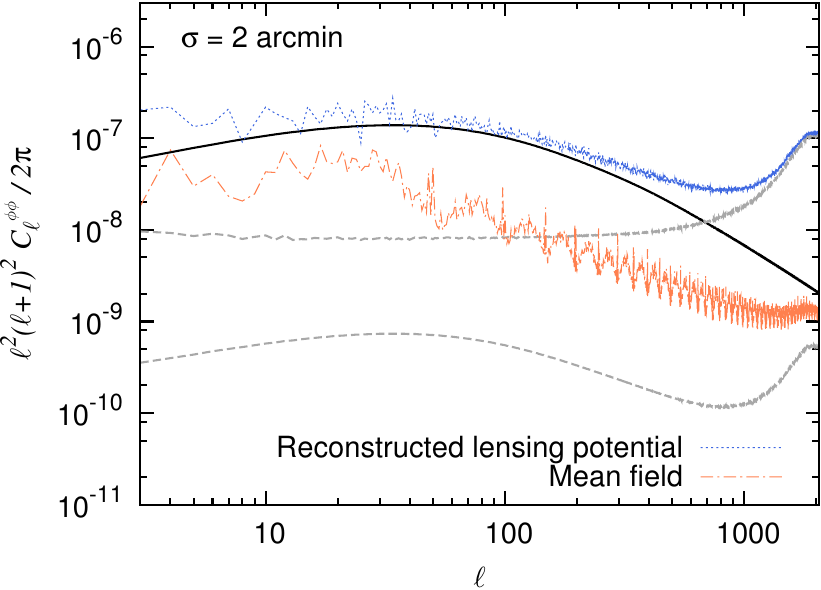}\label{clpp_mf_shift_l} }
\caption{Simulated angular power spectra for $\sigma = 2$arcmin subpatch center displacement. The line styles match the caption of Figure \ref{gain_s}.}
\label{shift_l}
\end{figure*}

Figures \ref{clbb_sys_shift_s} and \ref{clbb_sys_shift_l} show the spectra of the difference between the original CMB polarization map and the patchwork map where the subpatch centers have position errors. With displacements of a few arcminutes, the amplitudes of the differential spectra already respond approximately as the square of the displacement width (i.e., linear response in the map domain). The image on each subpatch is a polarization map located initially a small distance away. As the displacement width increases, the $E$-mode spectrum does not deviate much from the theoretical curve. In contrast, the $B$-mode spectrum clearly has an additional power derived from this systematic error. The displacements give the patchwork map a modulation similar to the cosmological gravitational lensing effect. Therefore, the bias in the $B$-mode spectrum becomes comparable to the amplitude of the lensing $B$-mode spectrum when the displacement width reaches a few arcminutes, comparable to the typical deflection angle of the gravitational lensing effect. Figures \ref{clpp_mf_shift_s} and \ref{clpp_mf_shift_l} show the result of the lensing potential reconstruction. As in the cases of gain error and polarization angle error, a sequence of spikes is observed in the mean-field spectrum.

To mitigate the impact of this systematic error, pointing accuracy is required to a high degree. In future observation projects, they aim to reduce their pointing error to be $\lesssim 10$ arcsec, e.g., Ref.  \cite{S42019}. If achieved, the impact on the lensing reconstruction analysis is not significant.

\subsection{Subpatch orientation rotation} \label{rec_rot}

\begin{figure*}
\hspace{-13pt}
\subfloat[]{
\includegraphics[width=3.05in]{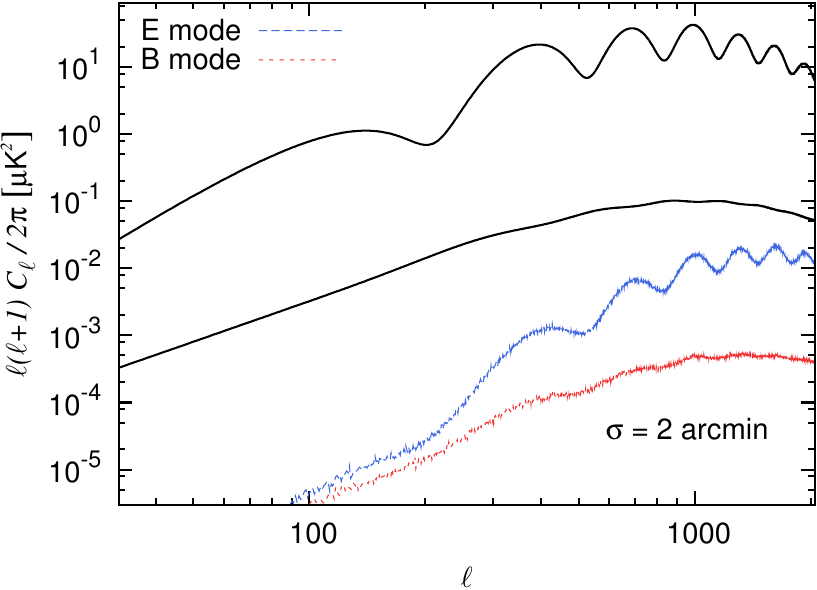}\label{clbb_sys_spin_s} }
\hspace{-4pt}
\subfloat[]{
\includegraphics[width=3.05in]{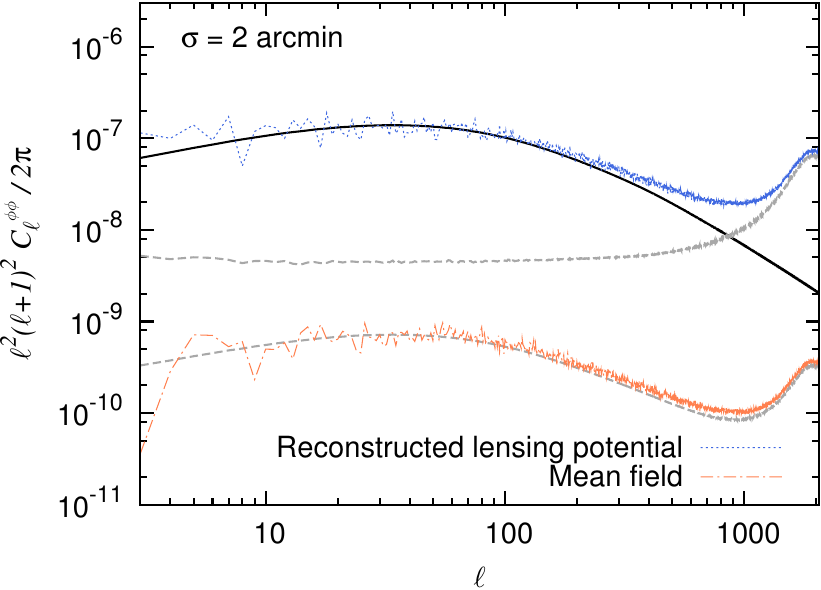}\label{clpp_mf_spin_s} }
\caption{Simulated angular power spectra for $\sigma = 2$arcmin subpatch orientation rotation. The line styles match the caption of Figure \ref{gain_s}.}
\label{spin_s}
\end{figure*}

\begin{figure*}
\hspace{-13pt}
\subfloat[]{
\includegraphics[width=3.05in]{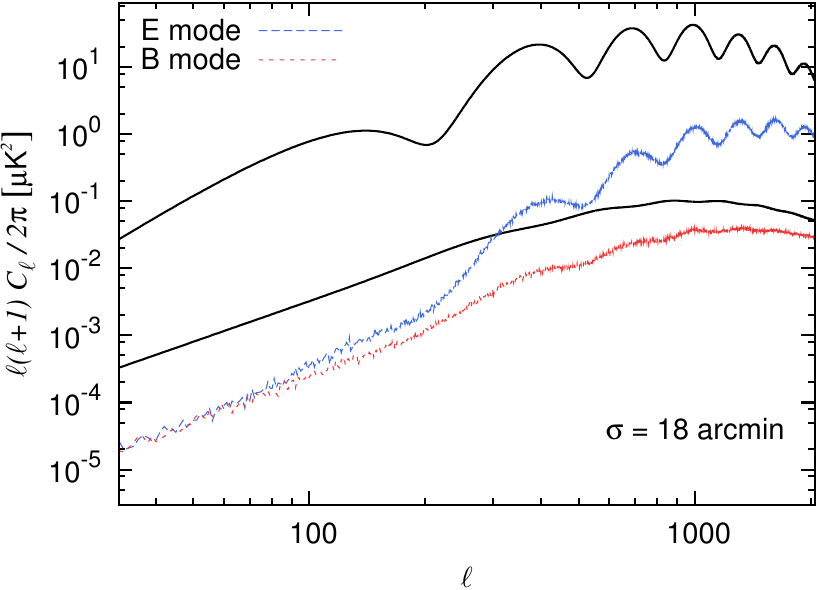}\label{clbb_sys_spin_l} }
\hspace{-4pt}
\subfloat[]{
\includegraphics[width=3.05in]{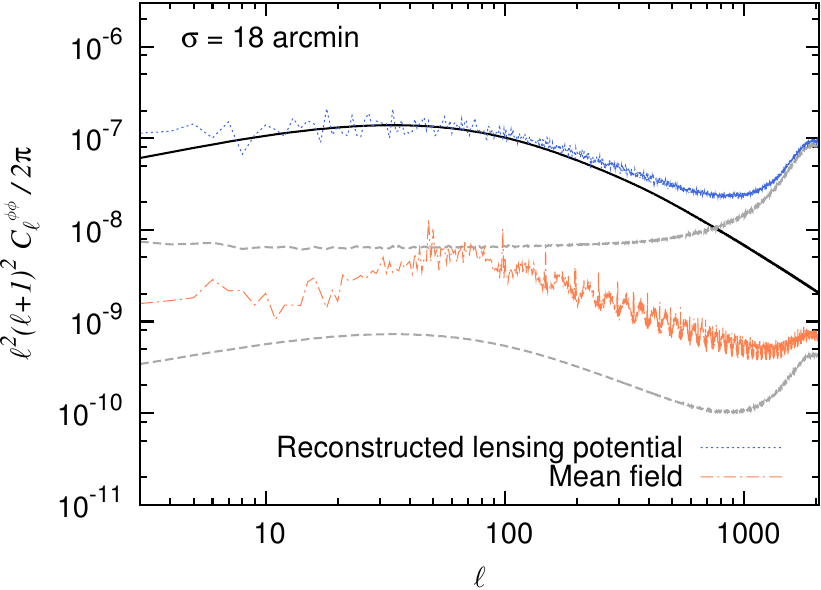}\label{clpp_mf_spin_l} }
\caption{Simulated angular power spectra for $\sigma = 18$arcmin subpatch orientation rotation. The line styles match the caption of Figure \ref{gain_s}.}
\label{spin_l}
\end{figure*}

Figures \ref{clbb_sys_spin_s} and \ref{clbb_sys_spin_l} show the spectra of the difference between the original CMB polarization map and the patchwork map where the subpatch orientations are slightly rotated. If the rotation is interpreted as positional misalignment at each sky point, the magnitude of the misalignment is equivalent to the rotation angle multiplied by the subpatch size ($\sim 1/10$ radian). Therefore, the apparent response of the spectra is about $100$ times smaller than in the case of subpatch center displacement. Considering this, we find that the behaviors, including the reconstruction results seen in Figures \ref{clpp_mf_spin_s} and \ref{clpp_mf_spin_l}, are generally similar to those in the case of subpatch center displacement.

\section{Delensing} \label{delensing}
In this section, we discuss the delensing analysis by utilizing the reconstructed lensing potential. By convolving the reconstructed lensing potential into the $E$ modes, we construct the delensing template as
\begin{equation}\label{B-temp}
\begin{split}
 \widehat{B}^{\rm lens}_{\ell m} &= -i \sum_{\ell'm'}\sum_{LM}
  \left(
   \begin{array}{ccc}
    \ell & \ell' & L \\
    m & m' & M \\
   \end{array}
  \right)
 S^{(-)}_{\ell\ell'L} W^{E}_{\ell'} W^{\phi}_L \widehat{E}_{\ell'm'}^{*}\widehat{\phi}_{LM}^{*} .
\end{split}
\end{equation}
Here, the Wiener filters are defined as follows:
\begin{equation}\label{Wiener}
\begin{split}
W^{\phi}_L &= \frac{ C^{\phi\phi}_L }{ C^{\phi\phi}_L+A_{L}^{EB} } , \\
W^{E}_{\ell} &= 1.
\end{split}
\end{equation}
The $E$ modes are not filtered because we are considering noiseless cases in this study. The delensing residual is defined as the difference between the original CMB $B$ modes (without systematic errors) and the template given by the above equation. Evaluating the ratio of the delensing residual spectrum ($\widehat{C}_{\ell}^{BB,{\rm res}}$) to the theoretical spectrum of the lensing $B$ modes, we call it delensing residual fraction. A small residual fraction means efficient delensing. The original $B$ modes to be delensed are assumed to be provided by large-scale CMB polarization data from satellite observations such as LiteBIRD \cite{Sekimoto2018,Lee2019}.

\begin{figure*}
\hspace{-13pt}
\subfloat[]{
\includegraphics[width=3.05in]{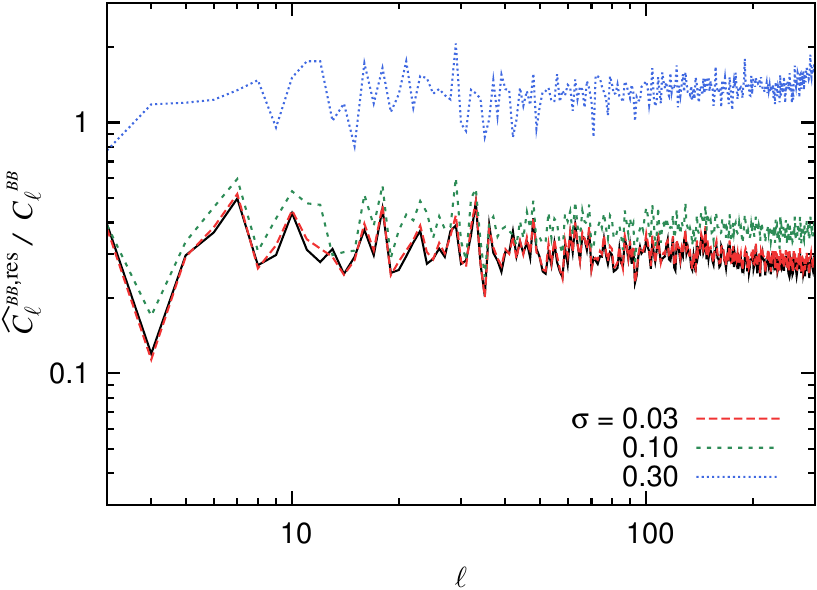}\label{clbb_del_gain_pol} }
\hspace{-4pt}
\subfloat[]{
\includegraphics[width=3.05in]{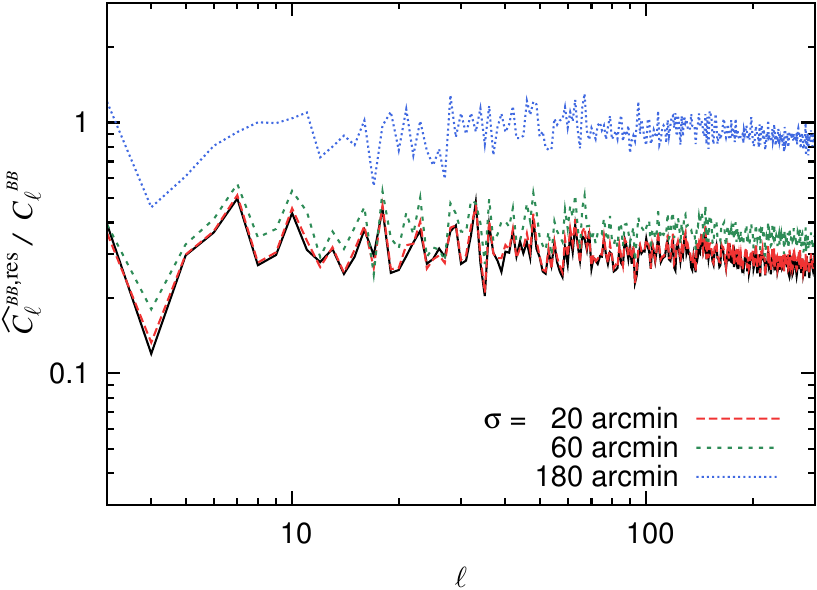}\label{clbb_del_angle_pol} }

\hspace{-13pt}
\subfloat[]{
\includegraphics[width=3.05in]{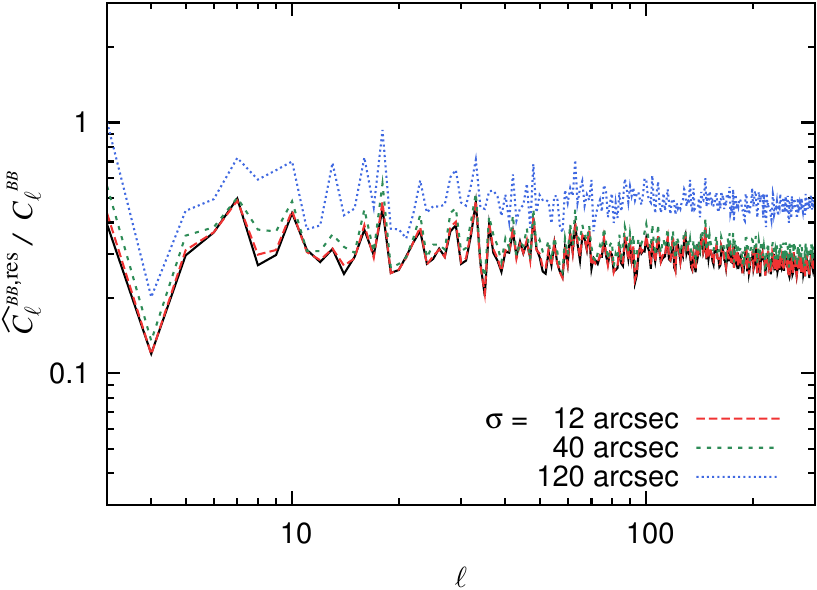}\label{clbb_del_shift_pol} }
\hspace{-4pt}
\subfloat[]{
\includegraphics[width=3.05in]{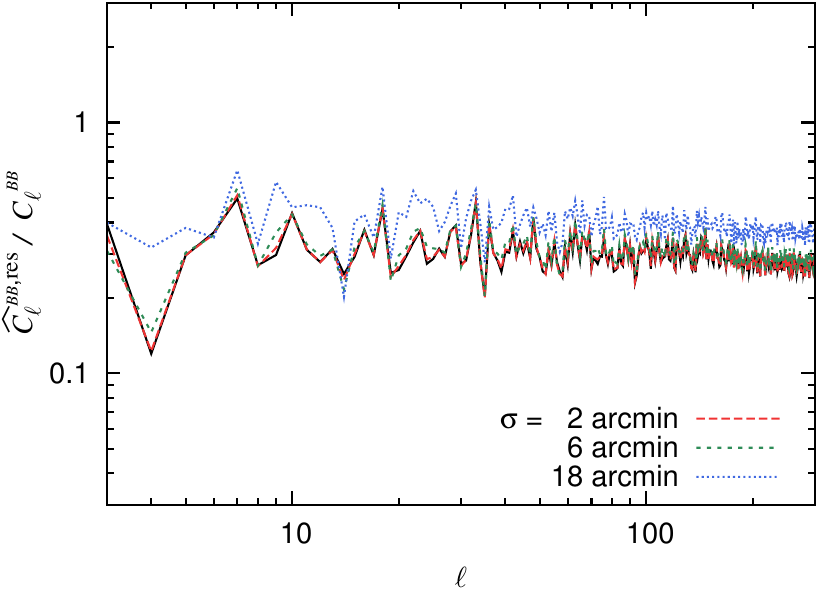}\label{clbb_del_spin_pol} }
\caption{
Dependence of delensing residual fraction on error amplitude. {\it Top-left}: residual gain error. {\it Top-right}: residual polarization angle error. {\it Bottom-left}: subpatch center displacement. {\it Bottom-right}: subpatch orientation rotation. The thick black line is for the case without any systematic errors.
}
\label{clbb_del_pol}
\end{figure*}

Even in the cases of subpatch misalignments, we evaluate the delensing residual from the difference between the original $B$ modes without the contribution from the systematic errors and the delensing template. This is different from the procedure once we adopted in Ref. \cite{SYSPTEP2021}. As a result of applying those systematic errors to the subpatches, the vector field of sky-points misalignment will contain both even-parity and odd-parity components to the same extent. Hence, the cancellation of the systematics-induced contributions by the delensing template is ineffective. More importantly, note that in the situation we have in mind, the patchwork polarization map is the source of information for lensing reconstruction, not the target of delensing.

Our delensing strategy involves three options, each with its own considerations. The first option is to use the lensing potential reconstructed from the patchwork polarization map for the convolution calculation. The second option is to create a patchwork map of lensing potentials, assuming that the lensing potentials are reconstructed for each subpatch, and then convolve it into the $E$ modes to evaluate the delensing template. However, this approach raises the question of whether an appropriate delensing template can be made without information on the large-scale lensing potential. The third method is to construct the delensing template for each subpatch. Unfortunately, this last approach is not viable for the delensing analysis in our scenario, as this delensing template cannot have information on scales larger than the subpatch size. Therefore, we adopt the first method, which uses the lensing potential reconstructed from the patchwork polarization map. The second method is discussed in more detail in Appendix \ref{phi_map}.

There are possible choices about $E$ modes with which we evaluate the delensing template. We use the $E$ modes based on the patchwork polarization map, assuming that ground-based observations provide high-resolution data that is desirable for the lensing analysis. Then, the information in $\ell < 50$ does not contribute to the delensing template, and we implement this by introducing a low ell cutoff to the Wiener filter in the above equation. Besides, the $E$ modes we use here are subjected to the systematic errors. Before proceeding, we would like to discuss what would happen if the original unmodulated CMB $E$ modes were used for the template construction. In the cases of gain error and polarization angle error, the delensing template is closer to the actual lensing $B$ modes because there is no systematics-induced contribution to the $E$ modes, and the delensing efficiency improves. On the other hand, in the cases of subpatch misalignments, the delensing efficiency degrades. The lensing potential reconstructed from such misaligned polarization maps can be interpreted to contain the same misalignments. It leads to a mismatch between the original CMB $E$ modes and the reconstructed lensing potential, resulting in a delensing template that is inaccurate on large scales. In our actual analysis, we adopt the $E$ modes based on the patchwork map. Although this delensing template also contains the same misalignments, the template's induced $B$ modes have little power on large scales. Since the delensing target is the large-scale lensing $B$ modes, they only slightly affect the delensing efficiency. However, this property should be considered when performing multi-tracer delensing in which other mass tracers, such as CIB and galaxies, are also used \cite{Sherwin:2015,Namikawa:2015:delens,LBDELENS}.

Each of the four panels in Figure \ref{clbb_del_pol} shows the response of the delensing efficiency to the magnitude of the systematic error concerned, i.e., gain error, polarization angle error, subpatch center misalignment, or subpatch orientation rotation. As discussed in Ref. \cite{SYSPTEP2021}, the systematic errors affect delensing mainly by propagating through the Gaussian bias in the reconstructed lensing potential. In each error case, the delensing efficiency quickly improves as the small-scale power given to the patchwork map by its applied error begins to be suppressed below the original CMB signals' power. As in the discussions of lensing reconstruction, by comparing these results to the calibration accuracies expected to be achieved in future observations, it can be concluded that the impact of those systematic errors on the delensing efficiency is insignificant. Furthermore, by examining the statistics of the samples generated by Monte Carlo simulations, we confirm that the variance of the residual power also decreases with the degree of suppression of the residual power itself. This means that the delensing method discussed here does indeed reduce uncertainty due to the lensing $B$ modes when analyzing data to search for the primordial gravitational waves.

\section{Summary and discussion} \label{summary}
In this paper, we have reported the impact of systematic errors on the lensing analysis based on a patchwork map of CMB polarization. The patchwork map consisted of small subpatches, and each constituent polarization map was associated with systematic errors that were coherent within its corresponding subpatch. Using numerical simulations, we investigated the effects from absolute calibration residuals of gain and polarization angle and those from subpatch misalignments. In each systematic error case, we discussed the behavior of the systematics-induced $E$ and $B$ modes in the patchwork map and the response of its reconstructed lensing potential and delensing efficiency.

As discussed in Section \ref{reconstruction}, large-scale Gaussian bias and mean field in the reconstructed lensing potential are well suppressed within calibration accuracies expected to be achievable in future observations. Therefore, the systematic errors argued in this work are not considered obstacles in evaluating multipole moments on scales larger than the subpatch size. In the delensing analysis, the effect of small-scale Gaussian bias is its primary issue. As shown in Section \ref{delensing}, it is feasible to mitigate the effect of the systematic errors on delensing sufficiently. It can be rephrased that, in practical situations, uncertainties due to noise and finite beam size determine the delensing efficiency. 

In this study, we did not include noise in our simulations. Therefore, baseline subtraction, usually performed to deal with $1/f$-type low-frequency noise, was also not included in the simulations. We performed another suite of simulations with a CMB polarization map, in which its power on scales larger than the subpatch size was artificially removed, and confirmed that the baseline subtraction does not change the conclusions of this study. A detailed discussion of the effects of the $1/f$ noise and baseline subtraction on the lensing analysis is found in our previous study \cite{Namikawa2014a}.

In the simulations of this study, the patchwork polarization map consists of the subpatches, which are uniformly about $7$ degrees in size. If the subpatch size is smaller, for a fixed value of error magnitude ($\sigma$ \footnote[5]{In the case of subpatch orientation rotation, replace with $\sigma$ multiplied by subpatch size.}), the power of the modulation-field spectrum, mentioned in Section \ref{mapsim_sys}, moves to smaller scales. In such a situation, gain error and subpatch misalignments give larger systematics-induced $E$ and $B$ modes, as suggested in Ref. \cite{SYSPTEP2021}. In other words, in the cases of those systematic errors, the smaller the subpatch size, the more stringent the requirement for error mitigation is expected to be. Roughly speaking, the power of the small-scale modulation field doubles about every time the patch size is halved, so it is estimated that $\sigma$ \footnotemark[5] needs to be suppressed by a factor of $\sim 1/\sqrt{2}$. In the case of polarization angle error, no such clear trend is found. On the other hand, in a situation where subpatches of various sizes coexist in the patchwork, its systematic error requirements are expected to be intermediate between the requirement corresponding to the largest of the subpatches and that to the smallest. Also, its mean-field spikes are not aligned neatly, and the power is distributed, resulting in a relatively smooth mean-field spectrum. Considering all these factors, larger subpatch sizes are still preferable. We look forward to efforts to develop instruments and analytical techniques that will extend the range of scales over which coherent observations can be made \cite{ABS2014,Hill2016,GB2020,Destripe2023}.

As long as we use CMB observations as our information source, excluding information on the Galactic plane (and nearby regions) is inevitable. It degrades the resolution of the reconstructed lensing potential in $\ell$-space. The degradation might manifest itself as correlations between multipole moments. There is also a mean field on scales beyond the size of the excluded region. Some perspective on these matters can be obtained from the discussions in Refs. \cite{BenoitLevy:2013,LBLENS}.

The sky coverage of CMB-S4 is envisioned to reach several dozen percent of the entire sky \cite{S42019}. Given such future observing trends, the prospect of lensing reconstruction from $\ell \sim O(1)$ using a patchwork polarization map is optimistic. The findings of this study are helpful in the initial stages of instrument design and observation planning. We are considering extending our work to simulation studies in accordance with specific observation setups.

\section*{Acknowledgment}

We acknowledge the use of {\tt Healpix} \cite{Gorski:2004by}, {\tt Lenspix} \cite{Lewis:2005apr}, and {\tt CAMB} \cite{Lewis:1999bs} for the simulations. This work is supported by JSPS KAKENHI Grant Number JP21K03610.


%\bibliographystyle{ptephy}
%\bibliography{sample}
%
% once the .bbl file has been generated then place the text in your article.

\bibliography{refs}

\newpage

\appendix

\section{Patchwork of lensing potential maps and delensing
} \label{phi_map}
In this appendix, we discuss the delensing analysis using a patchwork map of lensing potentials mentioned in Section \ref{delensing}. The patchwork map consists of the lensing potentials, each of which is separately reconstructed in its corresponding subpatch. We use the same patchwork polarization map for $E$ modes to be convolved as in the discussion in Section \ref{delensing}, treating their multipole moments with $\ell < 50$ as not contributing to the delensing template. (To add, the behavior of the delensing efficiency when using the original CMB $E$ modes is the same as in the case considered in Section \ref{delensing}.) We also assume that the multipole moments of the lensing potential in $\ell < 50$ do not contribute to the delensing template. This treatment is intended to reflect the situation where each lensing potential is reconstructed within its subpatch.

Let us first discuss the case of gain error. Observing equations \ref{EB-est} to \ref{EB-f} while noting that the polarization quantities are multiplied by a constant factor in each subpatch, we notice an optimistic scenario in which the lensing potential reconstructed for each subpatch is unaffected by the gain error. Specifically, by fitting the theoretical CMB spectra separately in each subpatch, we obtain an estimator that is formally identical to the one without the constant factor. Considering this property, we adopt the lensing potential map reconstructed in the absence of the gain error to construct the delensing template. (Note that the convolved $E$ modes contain the gain error.) Simulations based on such a setup show that the delensing efficiency is improved, e.g., for $\sigma=0.10$, the efficiency is comparable to the case without the gain error. However, the result depends on the feasibility of such an adjustment. Without it, we would not have achieved noticeable improvement. 

In the case of polarization angle error, there is no advantage in delensing using a patchwork map of lensing potentials. The degree to which the polarization angle error contaminates small-scale $E$ and $B$ modes shows little dependence on whether one analyzes each subpatch separately or analyzes their patchworked polarization map. There is no significant difference in the effect on the reconstructed lensing potential(s) and, therefore, in the delensing efficiency.

In the cases of subpatch misalignments, the misalignments do not affect the lensing analysis when achieved separately within each subpatch. To evaluate the delensing efficiency relevant to the situation, we first prepare an all-sky map of a lensing potential reconstructed under the assumption that there are no misalignments and then apply the misalignment operations to the lensing potential map. Using the resulting patchwork lensing potential map with its applied misalignments, we evaluate the delensing template. The change in the lensing potential caused by applying the misalignments can be interpreted as a mere perturbation to the lensing potential. Therefore, the effects from the misalignments are significantly reduced compared to the reconstruction from a patchwork polarization map. For example, in the case of $\sigma=120$arcsec subpatch center displacement, the Gaussian bias is suppressed to a degree comparable to the case without the displacement, and the delensing efficiency is also improved accordingly.

We have argued several possibilities above, some of which suggest that a patchwork of lensing potential maps reconstructed separately yields better delensing efficiency. However, practical situations and actual analysis procedures may override this advantage. We do not go deeper here and leave more detailed considerations for future work.

\end{document}